\documentclass[prd,floatfix,preprintnumbers]{revtex4}
\usepackage{amsmath}
\usepackage{graphicx,epsfig}
\usepackage{color}
\begin{document}
\title{Big Bang Nucleosynthesis with Stable $^8$Be and the Primordial Lithium Problem} 
\author {Richard T. Scherrer}
\affiliation{Department of Astronomy,
University of Illinois, Urbana, IL ~~61801}
\affiliation{and Department of Computer Science,
University of Illinois, Urbana, IL ~~61801}
\author {Robert J. Scherrer}
\affiliation{Department of Physics and Astronomy, Vanderbilt University,
Nashville, TN  ~~37235}

\begin{abstract}
A change in the fundamental constants of nature or plasma effects in the early universe could stabilize $^8$Be against decay into two $^4$He nuclei. 
Coc et al. examined the former effect on big bang nucleosynthesis as a function of $B_8$, the mass difference between two $^4$He nuclei and
a single $^8$Be nucleus, and found
no effects
for $B_8 \le 100$ keV.  Here we examine stable $^8$Be with larger $B_8$ and also allow for a variation in the
rate for $^4{\rm He} ~+~ ^4{\rm He} ~\longrightarrow~ ^8$Be to determine the threshold for
interesting effects.  We find no change to standard big bang nucleosynthesis
for $B_8 < 1$ MeV.  For $B_8 \gtrsim 1$ MeV and a sufficiently large reaction rate, a significant fraction of $^4$He is burned
into $^8$Be, which fissions back into $^4$He when $B_8$ assumes its present-day value,
leaving the primordial $^4$He abundance unchanged.  However, this sequestration
of $^4$He results in a decrease in the primordial $^7$Li abundance. Primordial abundances of $^7$Li
consistent with observationally-inferred values can be obtained for reaction rates similar
to those calculated for the present-day (unbound $^8$Be) case.
Even for the largest binding energies and largest reaction rates examined here, only a small fraction of
$^8$Be is burned into heavier elements, consistent with earlier studies.
There is no change in the predicted
deuterium abundance for any model we examined.
\end{abstract}

\maketitle

\section{Introduction}

Big bang nucleosynthesis (BBN) has long served as a useful constraint on the physics of the early universe.  In particular,
any change in the fundamental constants of nature could significantly alter BBN, allowing constraints to be placed on
such models \cite{KPW,dixit_sher,ss,Bergstrom,ichikawa,yoo,kneller,Dmitriev,Cyburt,LC,Coc,Dent,Berengut,8Be,Heffernan}.  For a review of BBN with time-varying constants, see Ref. \cite{Uzan}.

A particularly interesting possibility is that an appropriate change in the constants of nature might allow for the stability
of $^8$Be, which normally spontaneously fissions into $^4$He + $^4$He with a very short lifetime.
Coc et al. \cite{8Be} investigated the effects of stable $^8$Be on BBN.  More recently,
Adams and Grohs \cite{Adams} examined stellar evolution with stable $^8$Be.  Their goal was not to constrain such a model,
but rather to refute anthropic arguments for the fine-tuning needed to allow the triple-$\alpha$ reaction to proceed by showing
that stable $^8$Be could provide an acceptable alternative pathway for the production of heavier elements.
A completely different mechanism to stabilize $^8$Be has been suggested by Yao et al. \cite{Yao}. They proposed that plasma effects could stabilize $^8$Be in the early universe, obviating the need
for new physics.

To keep our results as general as possible, we will not assume a particular model for stable $^8$Be, but will instead treat
the $^8$Be binding energy as a free parameter.
Following Ref. \cite{8Be}, we define the mass difference between a single $^8$Be nucleus and two $^4$He nuclei to be
\begin{equation}
\label{B8def}
B_8 = 2 M( ^4 {\rm He}) - M( ^8{\rm Be}).
\end{equation}
Present-day measurements give $B_8 = -0.092$ MeV.  However, if the constants of nature during BBN were sufficiently different so as to make $B_8$ positive, then
$^8$Be would be stable, significantly altering the reaction network; a similar effect
might occur due to plasma effects in the early universe.
Coc et al. examined BBN for $B_8 \le 100$ keV and found no significant effect on any of the resulting element abundances.
Here, we extend this calculation to larger values of $B_8$.  Given uncertainties in the nuclear rates when
the constants of nature are allowed to change, we also parametrize the rate for $^4{\rm He} ~+~ ^4{\rm He} ~\longrightarrow~ ^8$Be
in terms of an overall multiplicative factor.  
In the next section, we present our calculations of the primordial element abundances and give the
results of these calculations.
We discuss our results in Sec. III.  We find that BBN can be significantly altered for $B_8 \sim 1-3$ MeV,
with a large reduction in the $^7$Li abundance, while the predicted deuterium and $^4$He abundances
are unchanged.  The nuclear fusion rates necessary to achieve this are similar to those calculated for
the present-day $^8$Be binding energy.

\section{Calculation of element abundances}
Consider first the standard model for BBN. (For recent reviews, see
Refs. \cite{Cyburt:2015mya,Mathews}). In the first stage of BBN, the weak interactions interconvert protons and neutrons, maintaining
a thermal equilibrium ratio:
\begin{eqnarray}\label{weak-interactions}
n~+~\nu _{e} &~\longleftrightarrow~ &p~+~e^{-},  \notag \\
n~+~e^{+} &~\longleftrightarrow~ &p~+~\bar{\nu}_{e},  \notag \\
n &~\longleftrightarrow~ &p+e^{-}~+~\bar{\nu}_{e},
\end{eqnarray}
while a thermal abundence of deuterium is maintained via
\begin{equation}
n ~+~ p ~\longleftrightarrow ~{\rm D} ~+~ \gamma.
\end{equation}
After the weak reactions drop out of thermal equilibrium at $T \sim 0.8$ MeV,
free neutron decay continues until
$T \sim 0.1$ MeV, when the thermal equilibrium abundance of deuterium becomes large enough
to allow rapid fusion into heavier elements.  Almost all of the remaining neutrons end up bound
into $^4$He, with a small fraction remaining behind in the form of deuterium. There is also some production
of $^7$Li via
\begin{eqnarray}
\label{Li7}
^4{\rm He} ~+~ ^3{\rm H}~ &\longrightarrow&~ ^7{\rm Li}~+~\gamma,\\
\label{Be7}
^4{\rm He} ~+~ ^3{\rm He}~ &\longrightarrow&~ ^7{\rm Be} ~+~ \gamma,
\end{eqnarray}
where the $^7$Be decays into $^7$Li via electron capture at the beginning of the recombination era \cite{Sunyaev}.

The element abundances produced in BBN depend on the baryon/photon ratio $\eta$, which can be independently
determined from the CMB.  We adopt a value of $\eta = 6.1 \times 10^{-10}$, 
consistent with recent results from Planck \cite{Ade}.  This value of $\eta$ yields predicted abundances
of D and $^4$He consistent with observations.  Recent observational estimates of D/H include
those of the Particle Data Group \cite{PDG}:  D/H $= (2.53 \pm 0.04) \times 10^{-5}$ and
Cooke {\it et al.} \cite{Cooke}: D/H $= (2.547 \pm 0.033) \times 10^{-5}$.  The primordial $^4$He abundance,
designated $Y_p$, is not as well established.  Izotov et al. \cite {Izotov} give
$Y_p = 0.2551 \pm 0.0022$, while Aver et al. \cite{Aver} give $Y_p = 0.2449 \pm 0.0040$. The Particle Data
Group limit is \cite{PDG} $Y_p = 0.2465 \pm 0.0097$. Given these discrepant estimates, a safe limit
on $^4$He is ${\rm Y}_p = 0.25 \pm 0.01$.  As noted, both the deuterium and $^4$He abundances are consistent with
the predictions of standard BBN with the CMB value for $\eta$.

The same cannot be said for the $^7$Li abundance.  The primordial lithium abundance is estimated to be
\cite{PDG}
\begin{equation}
^7{\rm Li/H} = (1.6 \pm 0.3) \times 10^{-10}.
\end{equation}
However, standard BBN with $\eta \sim 6 \times 10^{-10}$ predicts a primordial value for $^7$Li/H that is
roughly three times higher than this observationally-inferred value.  For this value of $\eta$, most of the primordial $^7$Li is produced
in the form of
$^7$Be, which decays into $^7$Li much later, as noted above.  This discrepancy between the predicted
and observationally-inferred primordial $^7$Li abundances has been dubbed the ``lithium problem," and it remains
unresolved at present (for a further discussion, see Ref. \cite{Fields}).  Hypothetical changes in the constants
of nature have been invoked previously as a possible solution of the lithium problem \cite{Coc}.

In standard BBN, $^8$Be is excluded from the reaction network, as it undergoes spontanous fission,
\begin{equation}
^8{\rm Be} ~\longrightarrow~ ^4{\rm He} ~+~ ^4{\rm He},
\end{equation}
with a lifetime $\sim 10^{-16}$ sec; the energy liberated in this fission is $-B_8$.  Here we assume that during the era of BBN, $B_8 > 0$, so that $^8$Be is stable.
Coc et al. \cite{8Be} examine a specific model for time variation of the fundamental constants, in which all of the binding
energies can be calculated as functions of the change in the nucleon-nucleon interactions. A similar
approach is taken by Epelbaum et al. \cite{Epelbaum}. Adams and Grohs \cite{Adams}
take a more general approach and discuss several ways in which changes in the fundamental constants
might alter $B_8$.  Since we are interested in isolating the particular effects of stable $^8{\rm Be}$, we shall adopt the latter
approach and treat $B_8$ as a free parameter.  The major limitation of our treatment is that we do not consider
changes in the other nuclear binding energies; these require the assumption of a specific
model like the one in Ref. \cite{8Be}.
We discuss this issue further in Sec. III.

Given the existence of stable $^8{\rm Be}$, the primary new reactions of importance are
\begin{equation}
\label{reaction1}
^4{\rm He} ~+~^4{\rm He} \longrightarrow ~^8{\rm Be} ~+~ \gamma,
\end{equation}
and
\begin{equation}
\label{reaction2}
^8{\rm Be} ~+~^4{\rm He} \longrightarrow ~^{12}{\rm C} ~+~ \gamma,
\end{equation}
along with the corresponding reverse reactions.
Estimated rates for these two reactions have been derived by Nomoto et al. \cite{Nomoto}, Langanke et al.
\cite{Langanke} and Descouvement and Baye \cite{DB}.  Adams and Grohs \cite{Adams} use the nonresonant reactions
from Ref. \cite{Nomoto}, while Coc et al. \cite{Coc} derived their own expressions for these rates based
on a particular model for changes in the nuclear interaction strength.  There are, of course,
large uncertainties in any calculation of this kind.  For example, the nonresonant rate calculated in Ref. \cite{Nomoto}
is not a direct-capture rate; instead, it represents the low-energy wing of the resonance at the current $^8$Be binding energy.
Any change in $B_8$ might have profound effects on this rate.  On the other hand, the calculation of Ref. \cite{8Be} assumes a particular
model for changes in the nuclear interaction strength.

We have chosen to parametrize the uncertainty in this calculation by expressing the rate for reaction (\ref{reaction1})
in terms of the standard expression for charged-particle interactions along with an overall multiplicative constant, which we allow to vary.
This allows us to the determine the threshold for interesting effects, which can then be compared (at least in order of magnitude)
to previous estimates for the rate.  For reaction (\ref{reaction2}), which is less important for our results, we follow Ref. \cite{Adams}
and use the nonresonant rate from Ref. \cite{Nomoto}.

Recall that for a charged particle reaction like reaction (\ref{reaction1}), the cross section as a function
of center-of-mass energy $E$ can be written as
\begin{equation}
\sigma(E) = S(E) E^{-1} \exp(-2\pi \eta),
\end{equation}
where $\eta$ is the Sommerfeld parameter, $\eta = Z_1 Z_2 e^2/\hbar v$, with $Z_1$ and $Z_2$ the charges
on the incoming nuclei and $v$ their relative velocities.  If the reaction is nonresonant, then $S(E)$ is generally
a slowly-varying function of $E$ (see, e.g., Ref. \cite{Clayton} for a pedagogical discussion).  The standard
procedure is to expand $S(E)$ in a power series around $E=0$ and convolve the cross-section with
the thermal distribution of nuclei.  For reaction (\ref{reaction1}) we obtain
an expression of the form \cite{wagoner}:
\begin{equation}
\label{rate0}
N_A \langle \alpha \alpha \rangle = T_9^{-2/3} ~\exp(-13.489~T_9^{-1/3}) \sum_{N=0}^5 F_N T_9^{N/3}~{\rm cm}^3~{\rm sec}^{-1}~{\rm mole}^{-1}.
\end{equation}
where $T_9$ is the temperature in units of $10^9$ K, and $N_A$ is Avogadro's number. 
In this expression for the reaction rate, the $F_N$ are functions of $S(E)$ and its first and
second derivatives at $E=0$ \cite{WFH}.
For the purposes of this study, we take only the constant term $F_0$ in Eq. (\ref{rate0}),
and we ignore all of the higher powers of $T_9^{1/3}$ so that
\begin{equation}
\label{rate1}
N_A \langle \alpha \alpha \rangle = F_0 T_9^{-2/3} ~\exp(-13.489~T_9^{-1/3})~{\rm cm}^3~{\rm sec}^{-1}~{\rm mole}^{-1}.
\end{equation}
Effectively, this amounts to treating $S(E)$ as a constant as $E \rightarrow 0$.
We do not claim that this is likely to be the most accurate
description of the form for the reaction rate.  However, it gives a simple one-parameter model for this rate
that can be compared (at least at the order-of-magnitude level) with other expressions for the cross section.
As noted earlier, for reaction (\ref{reaction2}), we simply use the nonresonant rate of Nomoto et al.
\cite{Nomoto}.  As we will see, this process has little impact on our final results.

The reverse reaction rates can be calculated from the forward rates using detailed balance.
For reactions of the form $i + j \longrightarrow k + \gamma$, we have (see, e.g., Ref. \cite{wagoner}),
\begin{equation}
\label{reverse}
\langle \sigma v \rangle_{k \gamma} = \frac{1}{1+ \delta_{ij}} \left (\frac{A_i A_j}{A_k}\right)^{3/2}
\frac{1.0 \times 10^{10} {\rm g/cm}^3}{\rho_B} T_9^{3/2} \exp(-11.605~ Q^*/T_9) \langle \sigma v \rangle_{ij},
\end{equation}
where we have
used the fact that all of the nuclei in reactions (\ref{reaction1}) and (\ref{reaction2}) are spin singlet states,
and $Q^*$ denotes the $Q$ values for reactions (\ref{reaction1})$-$(\ref{reaction2})
when $B_8$ is allowed to vary from its present value.
The present-day $Q$ values
for these two reactions are,
respectively, $Q_{\alpha \alpha} = -0.092$ MeV and $Q_{\alpha ^8{\rm Be}} = 7.27 ~{\rm MeV}$.  When
we allow the binding energy of $^8$Be to change, the new $Q$ values become
$Q^*_{\alpha \alpha} = B_8$ and $ Q^*_{\alpha ^8{\rm Be}}= 7.27 ~{\rm MeV} -
B_8$.

We expect reactions (\ref{reaction1}) and (\ref{reaction2}) to be the most important new pathways for the buildup of heavier
elements when $^8$Be is stable.  However, we have also examined the effects of the following reactions:
\begin{eqnarray}
\label{extrafirst}
^7{\rm Be} ~+~ n ~\longleftrightarrow~ ^8{\rm Be} ~+~ \gamma,\\
^7{\rm Li} ~+~ p ~\longleftrightarrow~ ^8{\rm Be} ~+~ \gamma,\\
^7{\rm Be} ~+~ ^2{\rm H} ~\longleftrightarrow~ ^8{\rm Be} ~+~ p,\\
^7{\rm Li} ~+~ ^2{\rm H} ~\longleftrightarrow~ ^8{\rm Be} ~+~ n,\\
^8{\rm Be} ~+~ n ~\longleftrightarrow~ ^9{\rm Be} ~+~ \gamma,\\
^8{\rm Li} ~+~ p ~\longleftrightarrow~ ^8{\rm Be} ~+~ n,\\
^8{\rm B} ~+~ n ~\longleftrightarrow~ ^8{\rm Be} ~+~ p,\\
^9{\rm Be} ~+~ p ~\longleftrightarrow~ ^8{\rm Be} ~+~ ^2{\rm H},\\
^{11}{\rm B} ~+~ p ~\longleftrightarrow~ ^8{\rm Be} ~+~ ^4{\rm He},\\
^{11}{\rm C} ~+~ n ~\longleftrightarrow~^8{\rm Be} ~+~ ^4{\rm He}.
\end{eqnarray}
To get a rough estimate of the effect of these reactions, we simply used the rates for the corresponding $2 \alpha$ reactions.
We
calculated the element abundances both with and without reactions (\ref{extrafirst})-(23).  Over our parameter
range of interest, we found no significant difference in the predicted element abundances when we included these additional
reactions, in agreement with the earlier results of Coc et al. \cite{8Be}.

We calculated the primordial element abundances using the AlterBBN computer code \cite{AlterBBN}, stripping out the triple-$\alpha$
reaction and replacing it with reactions (\ref{reaction1}) and (\ref{reaction2}).  We allowed
$B_8$ to vary up to 3 MeV, and we examined $F_0$ from $10^9$ to $10^{12}$.
Our results for $^4$He and $^8$Be are displayed in Fig. 1 for $F_0 = 10^{11}$, and Fig. 2
gives the $^7$Li abundance as a function of both $B_8$ and $F_0$.

Note first that the primordial $^2$H abundance (not displayed) is completely insensitive to $B_8$ even
for the largest values of $F_0$ we examined.  This makes
sense, as this abundance is determined by the rate of deuterium burning into
heavier elements, which is unaffected by helium burning into beryllium.  This implies that the
excellent agreement between the observed and predicted abundances of $^2$H is preserved (although
see the discussion in Sec. III regarding the deuterium binding energy).

We also find essentially no change
in any of the element abundances for small binding energies.  Our
results agree with Ref. \cite{8Be},
who found no significant change in the primordial element abundances for $B_8$ as large as 100 keV.  We can
extend this conclusion to larger values of the binding energy:  we find no discernable changes
in element abundances for $B_8$ as large as 600 keV, and significant changes only occur for $B_8 > 1$ MeV.

\begin{figure}[htb]
\centerline{\epsfxsize=6truein\epsffile{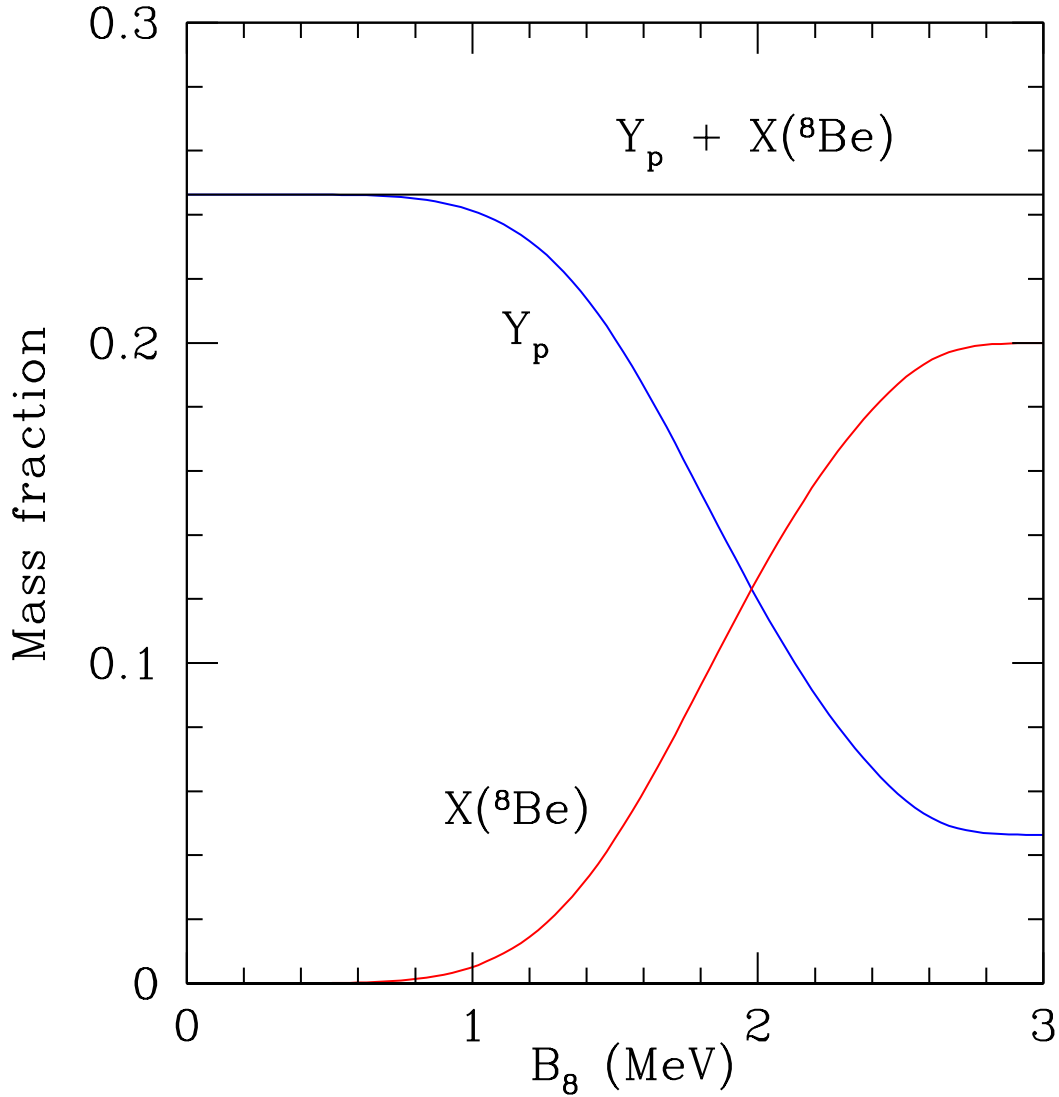}}
\caption{The primordial mass fractions of $^4$He (blue) and $^8$Be (red)
along with the sum of the $^4$He and $^8$Be mass fractions
(black) as
a function of $B_8$, the mass difference between two $^4$He nuclei and a
single $^8$Be nucleus, for $F_0 = 1.0 \times 10^{11}$, where $F_0$ parametrizes the $^4$He + $^4$He rate in Eq. (\ref{rate1}).}
\end{figure}

\begin{figure}[htb]
\centerline{\epsfxsize=6truein\epsffile{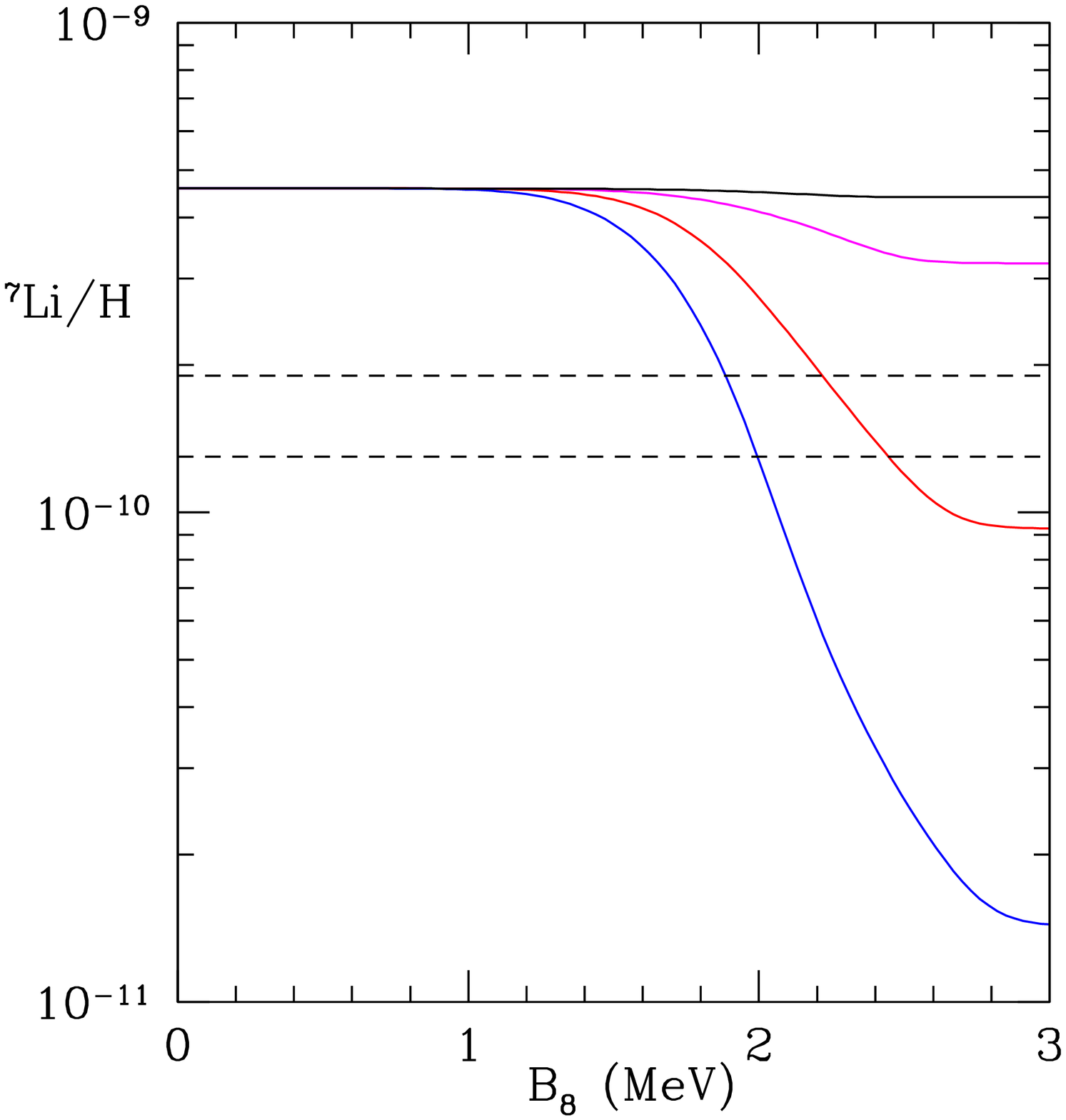}}
\caption{The abundance (relative to hydrogen) of $^7$Li, as a function of $B_8$, the mass difference between two $^4$He nuclei and a
single $^8$Be nucleus, for (top to bottom) $F_0 = 1.0 \times 10^9$ (black), $1.0 \times 10^{10}$
(magenta), $1.0 \times 10^{11}$ (red), $1.0 \times 10^{12}$ (blue), where $F_0$ parametrizes
the $^4$He + $^4$He rate in Eq. (\ref{rate1}).
The $^7$Li abundance is the sum of the primordial
$^7$Li and $^7$Be abundances, as the latter decays into the former.  Horizontal dashed lines give the range for the observationally
inferred value of $^7$Li/H.}
\end{figure}

In Fig. 1, we show $Y_p$ (the $^4$He mass fraction) and the
$^8$Be mass fraction as a function of $B_8$ for $F_0 = 10^{11}$.  The results for our other
values of $F_0$ are qualitatively similar.
As $B_8$ increases from 1 to 3 MeV, there is a sharp reduction in $Y_p$
and a corresponding increase in the $^8$Be abundance.
Naively, one might expect that this reduction in $Y_p$ to values far below the value estimated from observations
would rule out this model, but this is not the case.  We know that $B_8$ must assume its present-day negative
value at some time after
BBN.  When this occurs, $^8$Be will no longer be stable and will fission back into $^4$He.  Thus, the present-day
mass fraction of $^4$He will be given by the sum of the $^4$He and $^8$Be mass fractions.  We have plotted this sum
in Fig. 1.  Only an infinitesimal fraction of $^8$Be is burned into heavier nuclides, so this sum is constant and
equal to its value at $B_8 = 0$.  Thus, $^4$He, like $^2$H, is essentially unaffected by stable $^8$Be even for very
large binding energies and large rates for reaction (\ref{reaction1}).

However, large values of $B_8$ do have an important effect on the $^7$Li abundance, which
is displayed, relative to hydrogen, in Fig. 2.  The value of $^7$Li/H is clearly very sensitive
to both the $^8$Be binding energy and the rate of reaction (\ref{reaction1}).  For $F_0 \le 1.0 \times 10^9$,
there is essentially no effect on the lithium abundance.  Significant reduction begins to occur at 
$F_0 = 1.0 \times 10^{10}$ and $B_8 \ge 2$ MeV.  Lithium abundances in agreement with the observations
can be achieved as $F_0$ increases from $1.0 \times 10^{10}$ to $1.0 \times 10^{11}$, and for larger
$F_0$, there is a narrow range of values for $B_8$ for which
the predicted primordial lithium abundance agrees with the observationally-inferred value.
As in standard BBN, we find that most of the $^7$Li, at our chosen value of $\eta$,
is produced in the form of $^7$Be.  The physical
mechanism for this decrease is the sequestration
of $^4$He in the form of $^8$Be as seen in Fig. 1. This decrease in
the $^4$He abundance during BBN then inhibits reactions (\ref{Li7})-(\ref{Be7}).

The CNO elements are produced in very small amounts in standard BBN, with typical abundances
relative to hydrogen of CNO/H $\sim 10^{-15} - 10^{-14}$ \cite{CUV}.  A larger primordial production of
CNO elements would be interesting, as the results
of Ref. \cite{CC} suggest that the value of CNO/H begins
to affect the first generation of stars (population III) when CNO/H increases above $10^{-11}$.  However,
our results agree with those of Ref. \cite{8Be}; even for the largest $B_8$ and $F_0$ values we examined,
we see no significant primordial production of CNO elements.

\section{Discussion}

We find that BBN with stable $^8$Be can begin to produce interesting changes in the final element abundances for $B_8 \gtrsim 1$ MeV.
The deuterium and $^4$He abundances are unchanged, although the latter is sequestered in the form of $^8$Be until
$B_8$ drops below zero at late times.  This sequestration leads to a reduction in the $^7$Li abundance and can push it into
a regime consistent with observations for a sufficently large $^4$He + $^4$He rate.  We can compare the value of $F_0$
needed to produce this reduction in the lithium abundance with the nonresonant cross section of Ref. \cite{Nomoto}.
For $T_9 \sim 1 - 0.1$, the prefactor in Ref. \cite{Nomoto} corresponding to our $F_0$ lies between $4 \times 10^{11}$ and $2 \times 10^{10}$.
As we have already noted, it is not at all clear that the rates of Ref. \cite{Nomoto} can be extrapolated to a model with large
binding energies for $B_8$.  However, this comparison with Ref. \cite{Nomoto}
does indicate that the reaction rates examined here are not completely unreasonable.

A value of $B_8 \sim 1$ MeV is larger than has been considered in previous BBN calculations.  In the context
of plasma effects, it requires a very large Debye mass ($m_D \sim 3$ MeV) if one simply extrapolates the linear
approximation of Yao et al. \cite{Yao}.  This is larger than the value of $m_D$ predicted from plasma effects during BBN, although
there are some uncertainties in these calculations \cite{Yao}.  Such a large value of $B_8$ can be more plausibly achieved in the context
of time variation of the fundamental constants.  A value of $B_8 \sim$ 1 MeV can be obtained with
a change in the strong coupling constant of $\sim 15\%$, or changes in the quark masses or fine structure constant
by a similar amount \cite{Adams,Epelbaum}.

The major caveat in this discussion is that we have limited our analysis to changes in the $^8$Be binding energy alone.  This was intentional,
as we wished to isolate the effects of large changes in this binding energy in a model-independent way.  A realistic model would result
in changes to all of the nuclear binding energies, as in Ref. \cite{8Be}.  In changing the other nuclear binding energies,
the one likely to have the largest impact is deuterium \cite{yoo,kneller,Dmitriev,Coc}.  In the model presented in Ref. \cite{8Be},
$\sim$ 1 MeV values of $B_8$ would result in a $50\%$ increase in the deuterium binding energy.  A larger deuterium binding energy
would result in an earlier onset of nuclear fusion, leading to more $^7$Li, and potentially cancelling the reduction in $^7$Li noted here.
However, all of these conclusions depend on the particular model invoked to alter the nuclear binding energies.  A systematic estimate of the effects
of changing other binding energies can be found in Ref. \cite{Dent}.  It is possible that the plasma effects proposed by Yao et al. \cite{Yao}
would have a much larger effect on the $^8$Be binding energy than on the other nuclear binding energies, since these plasma effects
are sensitive to the existence
of the 92 keV resonance in $^8$Be. 

Our results indicate that it is
difficult to produce significant abundances of CNO elements in BBN even with MeV-scale binding energies for $^8$Be.
In that regard, the famous ``mass gap" at $A=8$
is misleading; the failure to produce heavier elements in the early universe is a result of the lower densities
and shorter times for nuclear fusion than prevail in stars \cite{8Be}.  This analysis ignores the
possibility that, for large values of $B_8$ and $F_0$, the build-up of a large mass fraction
of $^8$Be might allow the reaction $^8$Be + $^8$Be $\longrightarrow$ $^{16}$O + $\gamma$ to compete with reaction (\ref{reaction2}) as a mechanism for the production
of the CNO elements, but that seems unlikely in view of the large Coulomb barrier.  Of course, these results
are also sensitive to the assumed rate for $^8$Be + $^4$He; a rate that diverges from that of Ref. \cite{Nomoto}
could alter our conclusions regarding the CNO elements.

This work is admittedly speculative; our goal was to establish a threshold on the $^8$Be binding energy
and the $^4$He + $^4$He reaction rate that would produce a reduction in the primordial lithium abundance.  While
the possibility of solving the lithium problem through a change in the constants of nature,
including the binding energies of the light nuclei, is not new \cite{Coc},
the sequestration of $^4$He during BBN noted here represents a qualitatively new mechanism to achieve this.

\section{Acknowledgments}
We thank F.C. Adams, R. Galvez, and X. Yao for helpful discussions.


\begin{thebibliography}{99}

\bibitem{KPW}
E.W. Kolb, M.J. Perry, and T.P. Walker, \prd {\bf 33}, 869 (1986).

\bibitem{dixit_sher}
M. Sher and V.V. Dixit, \prd {\bf 37}, 1097 (1988).

\bibitem{ss}
R.J. Scherrer and D.N. Spergel, \prd {\bf 47}, 4774 (1993).

\bibitem{Bergstrom}
L. Bergstrom, S. Iguri, and H. Rubinstein,
\prd {\bf 60}, 045005 (1999).

\bibitem{ichikawa}
K. Ichikawa and M. Kawaskai, \prd {\bf 65}, 3511 (2002).

\bibitem{yoo}
J.J. Yoo and R.J. Scherrer, \prd {\bf 67}, 043517 (2003).

\bibitem{kneller}
J.P. Kneller and G.C. McLaughlin, \prd {\bf 68}, 103508 (2003).

\bibitem{Dmitriev}
V.F. Dmitriev, V.V. Flambaum, and J.K. Webb,
\prd {\bf 69}, 063506 (2004).

\bibitem{Cyburt}
R.H. Cyburt, B.D. Fields, K.A. Olive, and E. Skillman,
Astropart. Phys. {\bf 23}, 313 (2005).

\bibitem{LC}
B. Li and M.-C. Chu, \prd {\bf 73}, 025004 (2006).

\bibitem{Coc}
A. Coc, N.J. Nunes, K.A. Olive, J.P. Uzan, and E. Vangioni,
\prd {\bf 76}, 023511 (2007).

\bibitem{Dent}
T. Dent, S. Stern, and C. Wetterich,
\prd {\bf 76}, 063513 (2007).

\bibitem{Berengut}
J.C. Berengut, V.V. Flambaum, and V.F. Dmitriev,
Phys. Lett. B {\bf 683}, 114 (2010).

\bibitem{8Be}
A. Coc, P. Descouvemont, K.A. Olive, J.-P. Uzan, and E. Vangioni,
\prd {\bf 86}, 043529 (2012).

\bibitem{Heffernan}
M. Heffernan, P. Banerjee, and A. Walker-Loud,
[arXiv:1706.04991].

\bibitem{Uzan}
Uzan, J.-P., Living Reviews in Relativity {\bf 14}, 2 (2011).

\bibitem{Adams}
F.C. Adams and E. Grohs, Astropart. Phys. {\bf 87}, 40 (2017). 

\bibitem{Yao}
X. Yao, T. Mehen, and B. Muller, J. Phys. G: Nucl. Part. Phys. {\bf 43}, 07LT02 (2016).

\bibitem{Cyburt:2015mya}
R.H., Cyburt, B.D. Fields, K.A. Olive, and T.-H. Yeh,
Rev. Mod. Phys. {\bf 88}, 015004 (2016).

\bibitem{Mathews}
G.J. Mathews, M. Kusakabe, and T. Kajino, [arXiv:1706.03138].

\bibitem{Sunyaev}
R. Khatri and R.A. Sunyaev, Astron. Lett. {\bf 37}, 367 (2011).

\bibitem{Ade}
P.A.R. Ade, et al., Astron. Astrophys. {\bf 594}, A13 (2016).

\bibitem{PDG}
K.A. Olive {\it et al.} (Particle Data Group Collaboration),
Chin. Phys. C {\bf 38}, 090001 (2014).

\bibitem{Cooke}
R. Cooke, M. Pettini, K.M. Nollett, and R. Jorgenson,
Astrophys. J. {\bf 830} 148 (2016).

\bibitem{Izotov}
Y.I. Izotov, T.X. Thuan, and N.G. Guseva,
Mon. Not. R. Astron. Soc. {\bf 445}, 778 (2014).

\bibitem{Aver}
E. Aver, K.A. Olive, and E.D. Skillman, J. Cosmol. Astropart. Phys. {\bf 07}, 011,
(2015).

\bibitem{Fields}
B.D. Fields, Annu. Rev. Nucl. Part. Sci. {\bf 61}, 47 (2011).

\bibitem{Epelbaum}
E. Epelbaum, H. Krebs, T.A. Lahde, D. Lee, and U.-G. Meissner,
Eur. Phys. J. A {\bf 49}, 82 (2013).

\bibitem{Nomoto}
K. Nomoto, F.-K. Thielemann, and S. Miyaji, Astron. Astrophys. {\bf 149}, 239 (1985).

\bibitem{Langanke}
K. Langanke, M. Wiescher, and F.-K. Thielemann, Z. Phys. A {\bf 324}, 147 (1986).

\bibitem{DB}
P. Descouvemont and D. Baye, Phys. Rev. C {\bf 36}, 54 (1987).

\bibitem{Clayton}
D.D. Clayton, {\it Principles of Stellar Evolution and Nucleosynthesis}, (University of
Chicago Press, Chicago, 1983).

\bibitem{wagoner}
R.V. Wagoner, Astrophys. J. Suppl. {\bf 18}, 247 (1969).

\bibitem{WFH}
R.V. Wagoner, W.A. Fowler, and F. Hoyle, Astrophys. J. {\bf 148}, 3 (1967).

\bibitem{AlterBBN}
A. Arbey, Comput. Phys. Commun. {\bf 183}, 1822 (2012).

\bibitem{CUV}
A. Coc, J.-P. Uzan, and E. Vangioni
JCAP {\bf 10}, 050 (2014).

\bibitem{CC}
S. Cassisi and V. Castellani, Astrophys. J. S. {\bf 88}, 509 (1993).

\end{thebibliography}
\end{document}